\documentclass[aps,prb,twocolumn,superscriptaddress]{revtex4-1}

\usepackage{graphicx,amsmath}

\begin{document}


\title{Magnetic-field control of the electric polarization in BiMnO$_3$}


\author{I. V. Solovyev}
\email{SOLOVYEV.Igor@nims.go.jp}
\affiliation{
National Institute for Materials Science, 1-2-1 Sengen, Tsukuba,
Ibaraki 305-0047, Japan}
\author{Z. V. Pchelkina}
\affiliation{Institute of Metal Physics, Russian Academy of Sciences - Ural Division,
620041 Ekaterinburg GSP-170, Russia}


\date{\today}

\begin{abstract}
We present the microscopic theory of improper multiferroicity in BiMnO$_3$,
which can be summarized as follows: (1)
the ferroelectric polarization is driven by the hidden antiferromagnetic order
in the otherwise centrosymmetric $C2/c$ structure; (2) the relativistic spin-orbit interaction
is responsible for the
canted spin ferromagnetism. Our analysis is supported by numerical calculations
of electronic polarization
using Berry's phase formalism, which was applied to the low-energy model of BiMnO$_3$ derived
from the first-principles calculations.
We explicitly show how the electric polarization can be controlled by the
magnetic field and argue that BiMnO$_3$ is a rare and potentially
interesting material where ferroelectricity can indeed coexist and interplay
with the ferromagnetism.
\end{abstract}


\maketitle

\textit{Introduction}. Today, the term `multiferroics' is typically
understood in a broad sense, as the systems exhibiting
spontaneous
electric polarization and any type of magnetic ordering.\cite{Khomskii}
Such materials have a great potential for practical applications
in magnetic memories, logic, and magnetoelectric sensors,
and therefore attracted enormous attention recently.
Beside practical motivations, there is a
strong fundamental interest in
unveiling the microscopic mechanism of coupling between electric polarization
and
magnetic degrees
of freedom.
Nevertheless, the combination of ferroelectricity and \textit{ferro}magnetism,
what the term `multiferroicity' was originally introduced for, is rare.
Such a combination would, for example, provide an easy way for
manipulating the electric polarization ${\bf P}$ by the external magnetic field,
which
is directly coupled
to the net ferromagnetic (FM) moment, etc.
The canonical example, where spontaneous electric polarization was
believed to coexist with the FM ground state, is BiMnO$_3$.
However, the origin of such coexistence is largely unknown.
Originally, the ferroelectric (FE) behavior in BiMnO$_3$
was attributed to the highly distorted perovskite structure
stabilized by the
Bi$6s$ ``lone pairs''.\cite{SeshadriHill}
However, more resent experimental studies (Ref.~\onlinecite{belik_07})
and first-principles calculations (Ref.~\onlinecite{spaldin_07})
suggested that
the atomic displacements alone result in the centrosymmetric $C2/c$
structure, which is incompatible with the ferroelectricity.
In our previous papers (Refs.~\onlinecite{NJP08,JETP09}) we
put forward the idea
that the ferroelectricity in BiMnO$_3$ could be improper and associated
with some hidden antiferromagnetic (AFM) order. The purpose of this work
is to provide the complete quantitative explanation for the
appearance and behavior of the FE polarization in BiMnO$_3$.

\textit{Method}. The basic idea of our approach is to
construct an effective Hubbard-type model
\begin{equation}
\hat{\cal{H}}  =  \sum_{ij} \sum_{\alpha \beta}
t_{ij}^{\alpha \beta}\hat{c}^\dagger_{i\alpha}
\hat{c}^{\phantom{\dagger}}_{j\beta} +
  \frac{1}{2}
\sum_{i}  \sum_{\alpha \beta \gamma \delta} U_{\alpha \beta
\gamma \delta} \hat{c}^\dagger_{i\alpha} \hat{c}^\dagger_{i\gamma}
\hat{c}^{\phantom{\dagger}}_{i\beta}
\hat{c}^{\phantom{\dagger}}_{i\delta}
\label{eqn.ManyBodyH}
\end{equation}
for the
Mn$3d$-bands near the Fermi level and to
include the effect of all other (``inactive'') states
to the definition of
the model parameters of the Hamiltonian $\hat{\cal{H}}$. Thus, the model is constructed in the
basis of 40 Wannier functions in each unit cell
(including three $t_{2g}$- and two $e_g$-orbitals
for each spin and for each of the four Mn-sites), by
starting from the electronic
structure in the local-density approximation (LDA).
The Greek symbols denote the combination of
spin and orbital indices.
All parameters of $\hat{\cal{H}}$ are defined rigorously, on the basis of the
density functional theory (DFT).
The details can be found in the review article (Ref.~\onlinecite{review2008})
and in our previous papers (Refs.~\onlinecite{NJP08,JETP09}). Briefly,
the one-electron part ($t_{ij}^{\alpha \beta}$) is derived by using the generalized
downfolding method. One of important parameters in $t_{ij}^{\alpha \beta}$ 
is the large (about 1.5 eV) crystal-field splitting between two $e_g$-levels,
which is caused by the Jahn-Teller distortion and manifests itself in the
orbital ordering.
The screened Coulomb interactions
($U_{\alpha \beta \gamma \delta}$) are obtained by combining the
constrained DFT technique with the random-phase approximation (RPA):\cite{review2008}
namely, the screening by outer electrons
(such as $4sp$-electrons of transition metals)
and the change of spacial extension of the atomic wavefunctions
upon the change of occupation numbers
can be easily taken into account
by solving
Kohn-Sham equations
within constrained DFT approach.
On the other hand, the ``self-screening'' by the same type of electrons, which contribute
to other bands due to the hybridization effects
(for example, the $3d$-electrons in the oxygen band
will strongly screen the Coulomb interactions in the $3d$-band
near the Fermi level), is included in
the perturbative RPA treatment. The self-screening is very
important in solids and substantially
reduces the value of the effective Coulomb repulsion $U$
(defined as the screened Slater integral $F^0$)
in the
$3d$-band of manganites.\cite{JPSJ} In BiMnO$_3$, it
is only about 2.3 eV,\cite{NJP08} that
has important consequences on the behavior of interatomic
magnetic interactions.

  The model (\ref{eqn.ManyBodyH}) is solved
in the Hartree-Fock (HF) approximation:\cite{review2008}
$$
\left( \hat{t}_{\bf k} + \hat{\cal V} \right) |C_{n {\bf k}} \rangle =
\varepsilon_{n {\bf k}} |C_{n {\bf k}} \rangle,
$$
where $\hat{t}_{\bf k}$ is the Fourier image of $\hat{t}_{ij}$$=$$\| t_{ij}^{\alpha \beta} \|$
and, if necessary, includes the relativistic spin-orbit coupling (SOC),
$\hat{\cal V}$ is the self-consistent HF potential, and
$|C_{n {\bf k}} \rangle$ is the eigenvector
in the basis of Wannier functions
(where the spin indices are included in the definition of $n$).\cite{remark4}

  Once the orbital degeneracy is lifted by the strong lattice distortion,
the HF theory provides a good approximation for the ground-state properties.
The effect of correlation interactions, which can be treated as a perturbation to the HF solution,\cite{review2008}
on the magnetic ground state of manganites is partially compensated by the magnetic
polarization of the oxygen states:
if the former tend to stabilize AFM structures,
the latter favors the FM alignment.\cite{JPSJ} Due to this compensation,
the mean-field HF theory, formulated for the minimal $3d$-model, appears to be rather successful
for the ground state of manganites.

\textit{Magnetism and the inversion symmetry breaking}.
First, let us explain the main idea of our work.\cite{NJP08,JETP09}
What is the possible origin of multiferroic behavior of BiMnO$_3$ and how can it
be controlled by the magnetic field?

\begin{figure}
\begin{center}
\includegraphics[width=4cm]{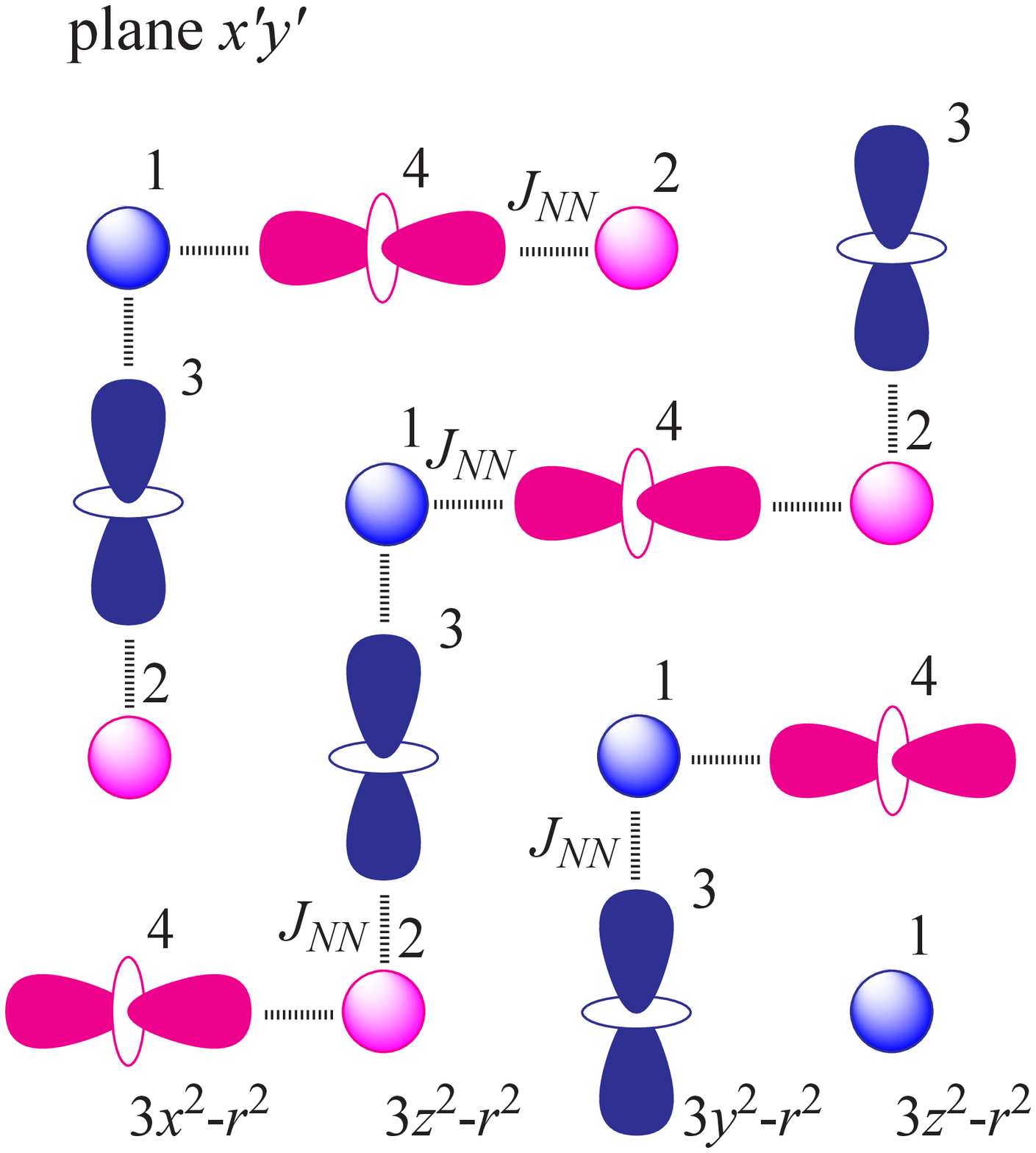} \ \
\includegraphics[width=4cm]{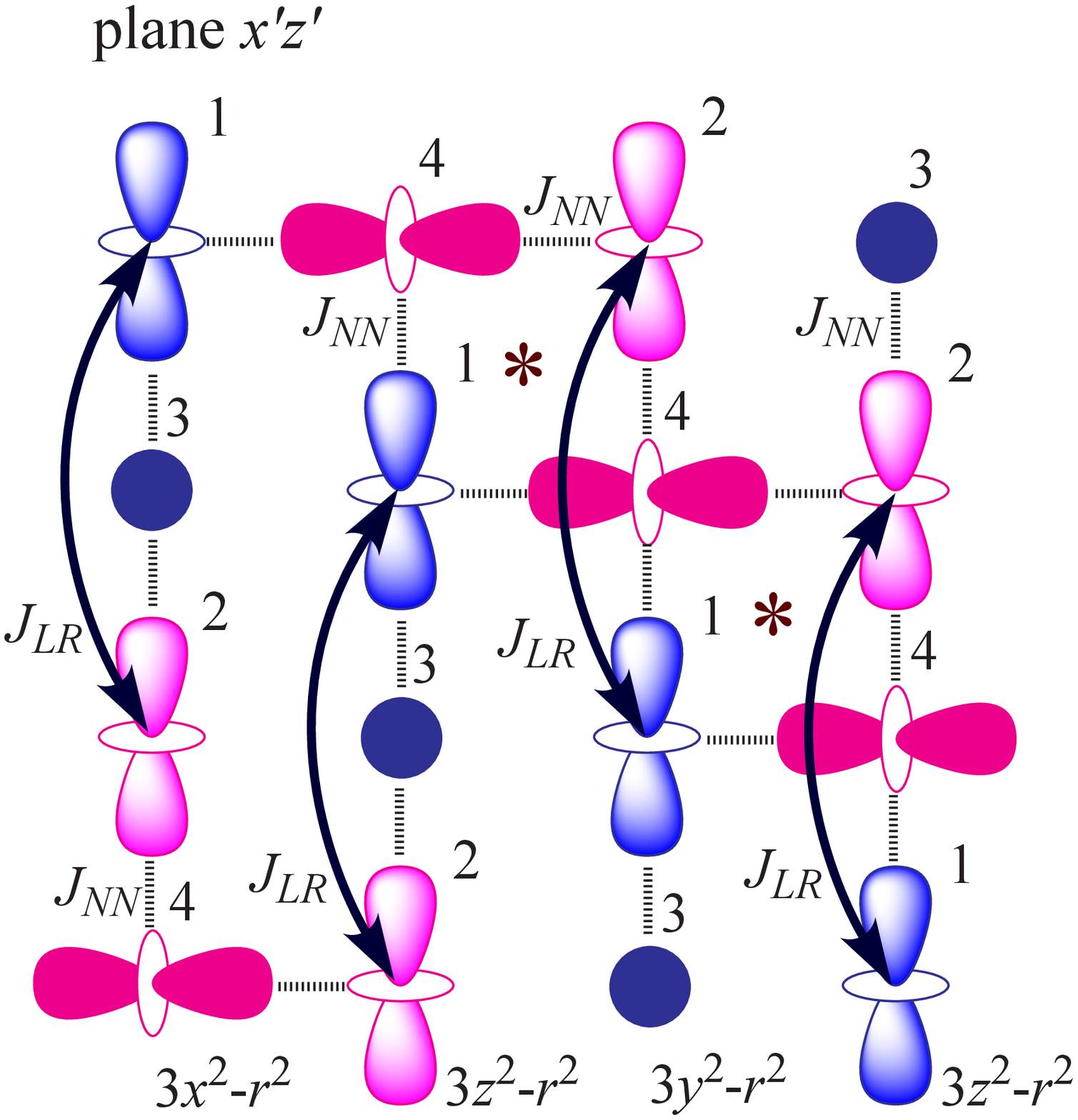}
\end{center}
\caption{\label{fig.OO}(Color online) Schematic view on the
orbital ordering and corresponding
interatomic magnetic interactions in the pseudocubic
$x'y'$ and $z'x'$ planes. In the unit cell of BiMnO$_3$,
there are four Mn sites (indicated by numbers), which form two inequivalent groups:
(1,2) and (3,4).
The nearest-neighbor FM interactions $J_{NN}$
operate in the hatched bonds. The atoms involved in the
long-range AFM interactions $J_{LR}$ are denoted
by arrows.
The inversion centers are marked by `$*$'. }
\end{figure}

(1) The lattice distortion leads the orbital ordering, which is
schematically shown in Fig.~\ref{fig.OO} in two pseudocubic planes (the orbital
ordering in the $y'z'$-plane is similar to the one in the $z'x'$-plane).
This orbital ordering predetermines the behavior of interatomic magnetic interactions,
which obey some general principles, applicable for manganites with both monoclinic ($C2/c$) and
orthorhombic ($Pnma$) structure,\cite{NJP08,JPSJ}
namely: besides conventional nearest-neighbor interactions
(shown by hatched lines), one can expect some
longer-range (LR) interactions between remote Mn-atoms, which operate via intermediate Mn-sites.
These sites are shown by arrows.

  (2) Why should the LR-interactions exist?
The answer is directly related to the fact that the on-site Coulomb repulsion $U$
is not particularly large. Therefore,
besides conventional superexchange (SE), there are other interactions, which
formally appear in the higher orders of the
$1/U$-expansion and connect more remote sites. This mechanism is rather similar to the SE
interaction via intermediate oxygen sites, but the role of the oxygen states is
played by the unoccupied $e_g$-orbitals of the intermediate Mn-sites.\cite{JPSJ} By mapping
HF total energies onto the Heisenberg model, one can obtain the following parameters
of
interatomic magnetic interactions:\cite{NJP08,remark2}  $J_{NN} \sim$ 5 and 6 meV
(where slightly different values correspond to different bonds) and $J_{LR} \sim$ $-$$3$ meV.
Thus, these interactions are at least comparable.
Besides them, there are finite (of the order $-$$1$ meV) interactions
in the bonds 1-2 and 4-4
across the inversion center, which
finally define the type of the magnetic
ground state of BiMnO$_3$.

  (3) Without SOC, the LR interactions tend to stabilize the AFM
$\uparrow \downarrow \downarrow \uparrow$ structure
(where the arrows denote the directions of spins for the four Mn-sites in the unit cell).
This AFM order destroys the inversion centers (shown by `$*$' in Fig.~\ref{fig.OO})
and thus should give rise to the FE polarization.
Since the $\uparrow \downarrow \downarrow \uparrow$ structure
satisfies the symmetry operation $\hat{T}$$\otimes$$\{ m_y|{\bf R}_3/2 \}$
(where $m_y$ is the mirror reflection $y \rightarrow -$$y$
associated with the one half of the
monoclinic translation ${\bf R}_3$, and $\hat{T}$ in the nonrelativistic case
flips the directions of spins, which are not affected by
$m_y$), ${\bf P}$ is expected to lie in the $zx$-plane.\cite{remark1}

\begin{figure}
\begin{center}
\includegraphics[width=6cm]{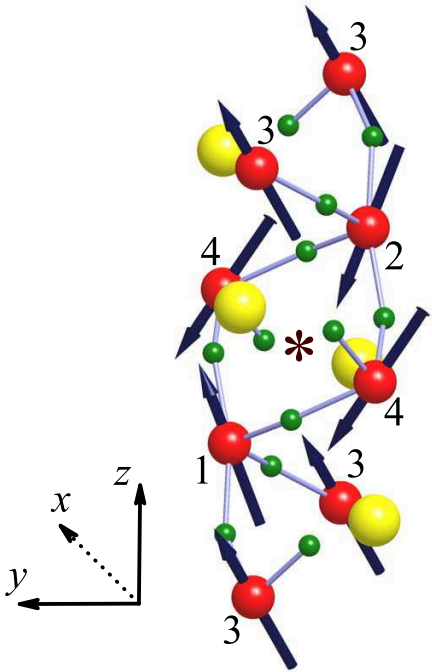}
\end{center}
\caption{\label{fig.MS}
(Color online) Fragment of crystal and magnetic structure corresponding to the
lowest HF energy. The Bi-atoms and indicated by the big light grey (yellow) spheres,
the Mn-atoms are indicated by the medium grey (red) spheres, and
the oxygen atoms
are indicated by the small grey (green) spheres. The directions of spin
magnetic moments are shown by arrows.
The inversion center is marked by the symbol `$*$'.
The left lower part of the figure explains the orientation of
the Cartesian coordinate frame.}
\end{figure}

  (4) Thus, the FE behavior in BiMnO$_3$
should be caused by the AFM order.
However, this conclusion seems to contradict to the FM ground state of
BiMnO$_3$.\cite{belik_07}
The contradiction can be reconciled by considering the relativistic SOC, which is
responsible for the weak ferromagnetism. Since the FM component is
additionally stabilized by the isotropic interactions $J_{NN}$, the ferromagnetism
is not so ``weak'', and the resulting magnetic structure,
obtained in the HF calculations for the low-energy model,
is strongly noncollinear (Fig.~\ref{fig.MS}).
It belongs to the space group $Cc$,
where the only nontrivial symmetry operation is $\{ m_y|{\bf R}_3/2 \}$ and
the magnetic moments in the relativistic case are transformed by $m_y$ as
auxiliary vectors.
Thus,
the net FM moment is aligned along the $y$-axis, while the $x$- and $z$-components
form the AFM structure. Other magnetic configurations
have higher energies.
The details can be found in Ref.~\onlinecite{JETP09}.

  By summarizing this part, the $C2/c$ symmetry of BiMnO$_3$ is spontaneously
broken by the hidden AFM order. The true magnetic ground-state of BiMnO$_3$
is strongly noncollinear, where the FM order along the $y$-axis coexists with the
AFM order, and related to it FE polarization, along the
$x$- and $z$-axes. Our scenario not only explains the coexistence of
ferroelectricity and ferromagnetism, but also shows how
the electric polarization ${\bf P}$ (and the symmetry of BiMnO$_3$)
can be controlled
by the external magnetic field ${\bf B}$$=$$(0,B_y,0)$ coupled to the FM moment.
This basic idea was formulated in Ref.~\onlinecite{JETP09}.
In the present work we are able to provide the numerical estimates for ${\bf P}$ and
to discuss its behavior in details.

  \textit{Electric polarization}.
Since the crystal structure of BiMnO$_3$ has the inversion symmetry,
there will be no ionic contribution to ${\bf P}$, and the main mechanism,
which will be considered below, is of purely electronic origin.
In principle, the magneto-elastic interactions in the
$\uparrow \downarrow \downarrow \uparrow$ structure may cause the
atomic displacements away from the centrosymmetric positions and
give rise to the ionic term. Nevertheless, such calculations would require the full
structure optimization, which cannot be easily incorporated in the
model analysis. The first-principles calculations for HoMnO$_3$ show
that electronic and ionic terms are at least comparable.\cite{Picozzi}
Thus, we expect that the electronic contribution alone could provide a good
semi-quantitative estimate for ${\bf P}$.
Moreover, the behavior of electronic contribution presents a fundamental interest
as it allows one to explain how ${\bf P}$ in improper multiferroics
is induced solely by the magnetic symmetry breaking.

  The modern theory of electric polarization allows one to relate
the change of ${\bf P}$
to the Berry phase of Bloch eigenstates.\cite{VKS,Resta1,Resta2}
It is particularly convenient to use the formulation by Resta,
where the Berry phase is computed on the discrete grid of ${\bf k}$-points,
generated by the $N_1$$\times$$N_2$$\times$$N_3$ divisions of the
reciprocal lattice vectors $\{ {\bf G}_a \}$.\cite{Resta2}
Then, the position of each point in the Brillouin zone is specified by three integer indices
($0 \leq s_a < N_a$):
$$
{\bf k}_{s_1,s_2,s_3} = \frac{s_1}{N_1}{\bf G}_1 + \frac{s_2}{N_2}{\bf G}_2 + \frac{s_3}{N_3}{\bf G}_3,
$$
and components of the electric polarization in the curvilinear coordinate frame
formed by ${\bf G}_1$, ${\bf G}_2$ and ${\bf G}_3$
can be obtained as\cite{Resta2}
\begin{equation}
\Delta P_a = - \frac{1}{V} \frac{N_a}{N_1 N_2 N_3}
\left[
\gamma_a(\infty) - \gamma_a(0)
\right],
\label{eqn:Pbasic}
\end{equation}
where $V$ is the unit cell volume,
$$
\gamma_1 = - \sum_{s_2 = 0}^{N_2-1} \sum_{s_3 = 0}^{N_3-1} {\rm Im} {\rm ln}
\prod_{s_1 = 0}^{N_1-1} {\rm det} S({\bf k}_{s_1,s_2,s_3},{\bf k}_{s_1+1,s_2,s_3}),
$$
and similar expressions hold for $\gamma_2$ and $\gamma_3$.
Eq. (\ref{eqn:Pbasic}) implies that the only meaningful quantity
in the bulk is the polarization difference between two states
that can be connected by
an adiabatic switching process.\cite{VKS,Resta1,Resta2}

  In the present case,
$S = \| \langle C_{n {\bf k}} | C_{n' {\bf k}'} \rangle \|$
is the overlap matrix, constructed from the
HF eigenvectors $|C_{n {\bf k}} \rangle$ in the occupied part of spectra,
taken
in two neighboring ${\bf k}$-points:
${\bf k}$$=$${\bf k}_{s_1,s_2,s_3}$ and ${\bf k}'$$=$${\bf k}_{s_1+1,s_2,s_3}$
for $\gamma_1$, etc.\cite{remark3}
The polarization (\ref{eqn:Pbasic}) was first computed in the
curvilinear coordinate frame and then transformed to the
cartesian frame shown in Fig.~\ref{fig.MS}.\cite{remark1}
In all the calculations, we used the mesh of $72$$\times$$72$$\times36$ points in the
Brillouin zone.

  Without SOC, the AFM alignment of spins at the sites 1 and 2 yields
finite polarization. However, the symmetry of the system also depends on the magnetic
configuration in the sublattice 3-4. As discussed  above,
the electric polarization in the
$\uparrow \downarrow \downarrow \uparrow$ structure lies in the $zx$-plane
($P_x = 2.1$ $\mu$C/cm$^2$ and $P_z = 0.1$ $\mu$C/cm$^2$).
The $\uparrow \downarrow \downarrow \uparrow$ structure can be transformed to the
$\uparrow \downarrow \uparrow \downarrow$ one with the same energy by the
symmetry operation $\{ C^2_y|{\bf R}_3/2 \}$ (where
$C^2_y$ is the $180^\circ$ rotation around the $y$-axis), which changes
the direction of ${\bf P}$: $P_{x(z)} \rightarrow -$$P_{x(z)}$.
On the other hand, the $\uparrow \downarrow \downarrow \downarrow$ structure
(which has higher energy)
is transformed to itself by $\{ C^2_y|{\bf R}_3/2 \}$,
and corresponding electric
polarization will be parallel to the $y$-axis ($P_y = 4.8$ $\mu$C/cm$^2$).
Other magnetic structures, characterized by the FM alignment of spins
at the sites 1 and 2 (such as $\uparrow \uparrow \uparrow \uparrow$ and
$\uparrow \uparrow \uparrow \downarrow$), preserve the inversion symmetry
and result in zero net polarization.

  Without SOC, one can easily evaluate separate contributions to ${\bf P}$ of
the states with the spin $\uparrow$ and $\downarrow$.
For the
$\uparrow \downarrow \downarrow \uparrow$ structure,
the vector of the electric polarization takes the following form:
${\bf P}^{\uparrow,\downarrow} = \frac{1}{2}(P_x,\pm P_y, P_z)$,
where
$P_y = 5.7$ $\mu$C/cm$^2$, and the values of $P_x$ and $P_y$ are listed above.
This result is very natural, because the distribution of the electron
density for each spin does not have any symmetry and, therefore,
the electric polarization ${\bf P}^{\uparrow,\downarrow}$ has all three
components. On the other hand, the electron density with the spin $\uparrow$
in the $\uparrow \downarrow \downarrow \uparrow$ AFM state can be transformed
to the one with the spin $\downarrow$ by the symmetry operation $\{ m_y|{\bf R}_3/2 \}$
and, therefore, $P^\uparrow_y = -$$P^\downarrow_y$.
Thus, in the total polarization ${\bf P} = {\bf P}^\uparrow + {\bf P}^\downarrow$,
the $x$- and $z$-components with different spins will sum up, while
the largest $y$-components will cancel each other.

  Furthermore, one can evaluate the individual contributions to ${\bf P}$ coming from the
$t_{2g}$-band,
which is separated by an energy gap from the $e_g$-band.\cite{NJP08}
This yields:
$P^{t_{2g}}_x = -$$0.8$ $\mu$C/cm$^2$ and $P^{t_{2g}}_z = -$$0.3$ $\mu$C/cm$^2$.
Thus, the $t_{2g}$-band is polarized \textit{opposite} to the $e_g$-band,
that substantially reduces the value of ${\bf P}$.

\begin{figure}
\begin{center}
\includegraphics[width=8cm]{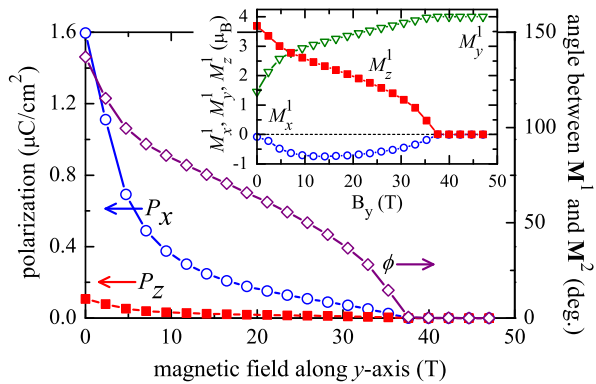}
\end{center}
\caption{\label{fig.field}(Color online) Magnetic-field dependence
of the electric polarization, the angle $\phi$ between spin magnetic
moments at the Mn-sites 1 and 2, and the vector of
magnetic moment at the site 1
(shown in the inset).}
\end{figure}

  The SOC results in the canting of spins away from the collinear
$\uparrow \downarrow \downarrow \uparrow$ state and towards the FM configuration.
It will \textit{reduce} the value of ${\bf P}$. In the HF ground-state
(see Fig.~\ref{fig.MS}), the angle $\phi$ between spin
magnetic moments at the sites 1 and 2 is 137$^\circ$, and the electric polarization is reduced till
$P_x = 1.6$ $\mu$C/cm$^2$ and $P_z = 0.1$ $\mu$C/cm$^2$.
This effect can be further controlled by the magnetic field, which is applied along the
$y$-axis and saturates the FM magnetization. Since the absolute value of the
local magnetic moment is nearly conserved, the increase of the FM component along
the $y$-axis will be compensated by the decrease of two AFM components
along the $x$- and $z$-axes.
The corresponding FE polarization will also decrease.
Results of HF calculations in the magnetic field are shown in Fig.~\ref{fig.field}.\cite{remark5}
Sufficiently large magnetic field
($\sim$35~Tesla) will align the magnetic
moments at the sites 1 and 2 ferromagnetically
($\phi$$=$$0$) and restore the $C2/c$ symmetry.\cite{JETP09}
The electric polarization follows the change of $\phi$
and complete disappears when $\phi$$=$$0$.
However, the decline of ${\bf P}$ is much steeper:
for example,
$P_x$ and $P_z$ are reduced by factor two already in the moderate field
$B_y \sim$ 5~Tesla, corresponding to $\phi \sim 100^\circ$.
Moreover, $P_z$ is always substantially smaller than $P_x$.

  \textit{Concluding remarks}.
We have proposed the microscopic theory of improper multiferroicity in BiMnO$_3$,
which is based on the inversion symmetry breaking by the hidden AFM order.
We have estimated the FE polarization and explicitly shown how it can
be controlled by the magnetic field. Our scenario still needs to be checked experimentally,
and apparently one important question here is how to separate the intrinsic
ferroelectricity in BiMnO$_3$ from extrinsic effects, caused by the defects.
For example, the values of the FE polarization obtained in the
present work, although comparable with those calculated for other improper
ferroelectrics on the basis of manganites,\cite{Picozzi} are substantially larger
than the experimental value 0.062 $\mu$C/cm$^2$ (at 87 K), which was
reported so far for BiMnO$_3$.\cite{dosSantos}
Nevertheless, we believe that systematic study of manganites with the monoclinic $C2/c$
symmetry and finding conditions, which would lead to the practical realization
of scenario proposed in our work, presents a very important direction, because
it gives a possibility for combining and intermanipulating
the \textit{ferro}electricity and \textit{ferro}magnetism in one sample.

  \textit{Acknowledgements}.
This work is partly supported by Grant-in-Aid for Scientific
Research (C) No. 20540337 from MEXT, Japan
and Russian Federal Agency for Science and Innovations, grant No. 02.740.11.0217.


\begin{thebibliography}{99}
\bibitem{remark3}
Strictly speaking, there will be two contributions to ${\bf P}$:
one is cause by the evalution of $|C_{n {\bf k}} \rangle$ in the
${\bf k}$-space
and the other one is the contribution of $\{ W^\alpha \}$ (the
Wannier basis of the low-energy model), which is expressed
in terms of the the matrix elements
$\langle W^\alpha | {\bf r} | W^\beta \rangle$.
Since the low-energy model
is constructed by starting from the nonmagnetic LDA band structure,
which preserves the parity of $\{ W^\alpha \}$, these matrix elements will vanish.
\bibitem{remark5}
The interaction term with the magnetic field is given by
$\hat{\cal H}_B = -$$\mu_B {\bf B} \cdot (2\hat{\bf s}$$+$$\hat{\bf l})$,
where $\hat{\bf s}$ and $\hat{\bf l}$ are the operators of spin and
orbital angular momentum, respectively.
\end{thebibliography}
\end{document}